\documentclass{PoS}

\usepackage{graphics,cite,amssymb,epsfig,float,graphicx}
\usepackage{amsmath, mathbbol}

\usepackage{amsfonts}

\def\be{\begin{equation}}
\def\ee{\end{equation}}
\def\bea{\begin{eqnarray}}
\def\eea{\end{eqnarray}}

\def\lsim{\;\raise0.3ex\hbox{$<$\kern-0.75em\raise-1.1ex\hbox{$\sim$}}\;}
\def\gsim{\;\raise0.3ex\hbox{$>$\kern-0.75em\raise-1.1ex\hbox{$\sim$}}\;}

\def\ben{\begin{enumerate}}  \def\een{\end{enumerate}}
\def\bit{\begin{itemize}}    \def\eit{\end{itemize}}
\def\beq{\begin{equation}}   \def\eeq{\end{equation}}
\def\ba{\begin{array}}       \def\ea{\end{array}}
\def\bea{\begin{eqnarray}}   \def\eea{\end{eqnarray}}

\renewcommand{\thetable}{\arabic{table}}

\title{Indirect searches for sterile neutrinos at a high-luminosity $\boldsymbol{Z}$-factory}

\ShortTitle{Indirect ~searches ~for ~sterile ~neutrinos ~at ~a ~high-luminosity~ $\boldsymbol{Z}$-factory}

\author{\speaker{V. DE ROMERI$^{1}$},  A. ABADA$^{2}$, S. MONTEIL$^{1}$,
  J. ORLOFF$^{1}$ and A.~M. TEIXEIRA$^{1}$ \thanks{PCCF RI 15-02, LPT Orsay 15-70}\\
  $^{1}$   Laboratoire de Physique Corpusculaire, CNRS/IN2P3 -- UMR 6533,\\ 
Campus des C\'ezeaux, 24 Av. des Landais, F-63177 Aubi\`ere Cedex, France\\
$^{2}$ Laboratoire de Physique Th\'eorique, CNRS -- UMR 8627, \\
Universit\'e de Paris-Sud, F-91405 Orsay Cedex, France\\
        E-mail: \email{deromeri@clermont.in2p3.fr}}

\abstract{A future high-luminosity $Z$-factory has the potential to investigate lepton flavour violation.
Rare decays such as $Z \to \ell_1^\mp \ell_2^\pm$
can be complementary to low-energy 
(high-intensity) observables of lepton flavour violation.
Here we consider two extensions of the Standard Model which add to its particle content one or more sterile neutrinos. We address the impact of the sterile fermions 
on lepton flavour violating $Z$ decays, focusing on potential searches
at FCC-ee (TLEP), and taking into account experimental and
observational constraints. 
We show that sterile neutrinos can 
give rise to contributions to BR($Z \to \ell_1^\mp \ell_2^\pm$) 
within reach of the FCC-ee. We discuss the complementarity between 
a high-luminosity $Z$-factory and low-energy charged lepton flavour
violation facilities.}

\FullConference{The European Physical Society Conference on High Energy Physics\\
		22--29 July 2015\\
		Vienna, Austria}

\begin{document}
\vskip -2.5cm
\section{Introduction}
\vskip -0.2cm
\noindent
Several extensions of the Standard Model (SM) add sterile neutrinos to the particle content in order to account for neutrino masses and mixings. These models are further motivated by anomalous 
(oscillation) experimental results,
as well as by certain 
indications from cosmology (see \cite{Kusenko:2009up,Abazajian:2012ys} and references therein). 
The existence of these sterile states may be investigated at colliders: 
for instance, the case for a
high luminosity circular $e^+e^-$ collider (called FCC-ee), operating at
centre-of-mass energies ranging from the $Z$ pole up to the top quark
pair threshold is being actively
studied~\cite{Gomez-Ceballos:2013zzn}.  Its characteristics 
should allow to obtain a typical peak luminosity at
the $Z$ pole of $\sim 10^{36} \rm cm^{-2} \rm s^{-1}$. A year of operation at the $Z$ pole
centre-of-mass energy would yield  $\sim 10^{12}$ $Z$ boson
decays to be recorded.
Motivated by the design study for such a powerful machine, we investigate
the prospects for searches for sterile neutrinos by means of rare {\emph {charged}} lepton flavour violating (cLFV) $Z$ decays \cite{Abada:2014cca}.
\vskip -1cm
\section{Leptonic $\boldsymbol{Z}$ decays in the presence of sterile
  neutrinos}\label{sec:Zdecays:gen} 
  \vskip -0.2cm
  \noindent
Lepton-flavour changing $Z$ decays are forbidden in the SM
due to the GIM mechanism~\cite{Glashow:1970st}, and their rates remain extremely small ($ \sim 10^{-54} - 10^{-60}$) even 
when lepton mixing is introduced.  The observation of such a rare decay would therefore serve as an indisputable evidence
of new physics~\cite{LFVZdecays,Perez:2003ad}. We consider here two extensions of the SM which introduce sterile Majorana fermions. The mixing in the neutral lepton sector
induced by these states 
also opens the possibility for flavour violation in
$Z\nu_i\nu_j$ interactions (flavour-changing neutral
currents), coupling both the left- and right- handed components of
the neutral fermions to the $Z$ boson. Together with the
charged-current LFV couplings, 
these interactions will induce an 
effective cLFV
vertex $Z \ell_1^\mp \ell_2^\pm$. 

\paragraph{Inverse Seesaw}

The Inverse Seesaw (ISS) mechanism~\cite{Mohapatra:1986bd} 
is an example of low-scale seesaw realisation which in full generality calls upon the introduction of at least two generations 
of SM singlets. Here, we consider the addition of three generations of right-handed (RH) neutrinos $\nu_R$ 
and of extra $SU(2)$ singlets fermions $X$, to the SM particle content.  Both $\nu_R$ and $X$
carry lepton number $L=+1$.
The ISS Lagrangian reads
$
\mathcal{L}_\text{ISS} \,=\, 
\mathcal{L}_\text{SM} - Y^{\nu}_{ij}\, \bar{\nu}_{R i} \,\tilde{H}^\dagger  \,L_j 
- {M_R}_{ij} \, \bar{\nu}_{R i}\, X_j - 
\frac{1}{2} {\mu_X}_{ij} \,\bar{X}^c_i \,X_j + \, \text{h.c.}\,,
$
where $i,j = 1,2,3$ are generation indices and $\tilde{H} = i \sigma_2
H^*$. Lepton number $U(1)_L$ is broken
only by the non-zero Majorana mass term $\mu_{X}$, while the Dirac-type 
RH neutrino mass term $M_{R}$ does conserve lepton number.
In the $(\nu_L,{\nu^c_R},X)^T$ basis, and after the electroweak symmetry breaking, 
the (symmetric) $9\times9$ neutrino mass matrix 
$\mathcal{M}$ is given by
\vskip -0.2cm
\begin{eqnarray}
{\cal M}&=&\left(
\begin{array}{ccc}
0 & m^{T}_D & 0 \\
m_D & 0 & M_R \\
0 & M^{T}_R & \mu_X \\
\end{array}\right) \, ,
\label{eq:ISS:M9}
\end{eqnarray}
with $m_D= Y^\nu \rm v$ the Dirac mass term,  
$\rm v$ being the vacuum expectation value of the SM Higgs boson.  
Under the assumption that $\mu_X \ll m_D \ll M_R$, the
diagonalization of ${\cal M}$ leads to an effective Majorana mass
matrix for the active (light) neutrinos~\cite{GonzalezGarcia:1988rw},
$ m_\nu \,\simeq \,{m_D^T\, M_R^{T}}^{-1} \,\mu_X \,M_R^{-1}\, m_D \,$.
The remaining six (mostly) sterile states form nearly degenerate pseudo-Dirac
pairs. 
The possibility of having sizeable mixings between the active and sterile states, 
will have a non-negligible impact for several observables, thus rendering the ISS framework phenomenologically appealing.

\vskip -0.25cm
\paragraph{The effective ``3+1 model''}\label{sec:3+1}
A simpler approach to address the impact of sterile fermions on rare cLFV $Z$ decays  consists in considering a minimal model
where only one sterile Majorana state
is added to the three light active neutrinos of the SM. 
This allows for a generic
evaluation of the impact of the sterile fermions for these processes.
In this simple toy model, no assumption is made on the underlying mechanism of
neutrino mass generation. The addition of an extra neutral fermion to the particle content 
translates into extra degrees of freedom:  the mass of the new sterile state $m_4$, three
active-sterile mixing angles $\theta_{i4}$, three new CP phases (two Dirac and one Majorana). 

\noindent
In our analysis, and for both hierarchies of the light neutrino spectrum, we
scan over the following range for the sterile neutrino mass: 
$
10^{-9} \text{ GeV } \lsim \, m_4 \, \lsim 10^{6} \text{ GeV},
$
while the active-sterile mixing angles are randomly varied in the
interval $[0, 2 \pi]$ \footnote{We always ensure that the the perturbative unitary
bound on the sterile masses and
their couplings to the active 
states is respected.}. 
All CP phases are also taken into account, and
likewise randomly varied between 0 and $2 \pi$. 

\vskip -0.25cm
\section{Constraints on sterile neutrino extensions of the
  SM}\label{sec:constraints}
\noindent
The introduction of sterile fermion states, which
have a non-vanishing mixing to the active neutrinos, leads to a modification of
the leptonic charged current Lagrangian:
\vskip -0.2cm
\begin{equation}\label{eq:cc-lag}
- \mathcal{L}_\text{cc} = \frac{g}{\sqrt{2}} {\bf U}^{ji} 
\bar{\ell}_j \gamma^\mu P_L \nu_i  W_\mu^- + \, \text{c.c.}\,,
\end{equation}
where ${\bf U}$ is the leptonic mixing matrix,
$i = 1, \dots, n_\nu$ denotes the physical neutrino states
and $j = 1, \dots, 3$ the flavour of the charged leptons. 
In the standard case of three neutrino generations,  ${\bf U}$ corresponds
to  the unitary matrix $U_\text{PMNS}$.
For $n_\nu >3$, the 
mixing between the left-handed leptons, which we denote by $\tilde U_\text{PMNS}$, 
corresponds to a $3 \times 3$ sub-block of ${\bf U}$, which can show some deviations from unitarity.
One can parametrise~\cite{FernandezMartinez:2007ms} the 
$\tilde U_\text{PMNS}$ mixing matrix as
$
U_\text{PMNS} \, \to \, \tilde U_\text{PMNS} \, = \,(\mathbb{1} - \eta)\, 
U_\text{PMNS},$
where the matrix $\eta$ encodes the deviation of the $\tilde
U_\text{PMNS}$ from unitarity~\cite{Schechter:1980gr,Gronau:1984ct}, 
due to the presence of extra neutral fermion states.
One can also introduce the invariant quantity
$\tilde \eta$, defined as
$\tilde \eta = 1 - |\text{Det}(\tilde U_\text{PMNS})|$,
particularly useful to illustrate the effect of the new
active-sterile mixings (corresponding to a deviation from unitarity of
the $\tilde U_\text{PMNS}$) on several observables.\\
The deviation from unitarity of ${\bf U}$ will induce a departure from the SM expected values of several observables.
In turn, this is translated into a vast array of constraints which we will apply to our analysis (see details and references in \cite{Abada:2014cca}).
Firstly, one has to ensure that the SM extension complies with 
{\it $\nu$-oscillation data}: we require compatibility with the corresponding best-fit 
intervals~\cite{Forero:2014bxa} (no constraints being imposed on the
yet undetermined value of the CP violating Dirac phase $\delta$).
We also apply {\it unitarity bounds} 
on the (non-unitary) matrix $\eta$;
these arise from non-standard neutrino interactions with matter,
and have been derived in~\cite{Antusch:2008tz,Antusch:2014woa}  
by means of an effective theory approach (valid for sterile masses
above the GeV, but  
below the electroweak scale, $\Lambda_\text{EW}$).
We further take into account {\it electroweak precision observables}
requiring, for instance, compatibility with LEP results
on $\Gamma(Z \to \nu \nu)$. 
LHC data on {\it invisible Higgs} decays already allows to constrain
regimes where the sterile states are below the Higgs mass. 
Negative {\it laboratory searches} for monochromatic lines in the
spectrum of muons from  $\pi^\pm \to \mu^\pm \nu$
decays 
also impose robust bounds on 
sterile neutrino masses in the MeV-GeV range. 
The introduction of singlet neutrinos with Majorana masses allows for 
new processes like lepton number violating interactions, among which 
{\it neutrinoless double beta decay} remains the most important
one. 
In our analysis, we evaluate the contributions of the sterile states
to the effective mass $m_{ee}$; 
we use 
the most recent constraint from EXO-200 (concerning future sensitivities we take
$|m_{ee}| \lesssim 0.01$ eV).
Further constraints arise from {\it leptonic and semileptonic decays of
pseudoscalar mesons}
$\ K,\ D, \ D_s$, $B$. 
Recent studies suggest that in the framework of the SM extended by
sterile neutrinos the 
most severe bounds arise from the violation of lepton universality
in leptonic kaon decays. 
Other than the rare decays occurring in the presence of nuclei, 
the new states can contribute to several 
{\it charged lepton flavour violating processes} such as  
$\ell \to \ell^\prime \gamma$, $\ell\to \ell_1\ell_1\ell_2$.
In our analysis we compute the contribution of the sterile states to
all these
observables 
imposing compatibility with the current experimental bounds.
Finally, a number of
{\it cosmological observations}~\cite{Kusenko:2009up}
put severe constraints on sterile neutrinos with a mass below the TeV. 
\section{Results}
\begin{figure}[hbt!]
\begin{tabular}{cc}
\epsfig{file=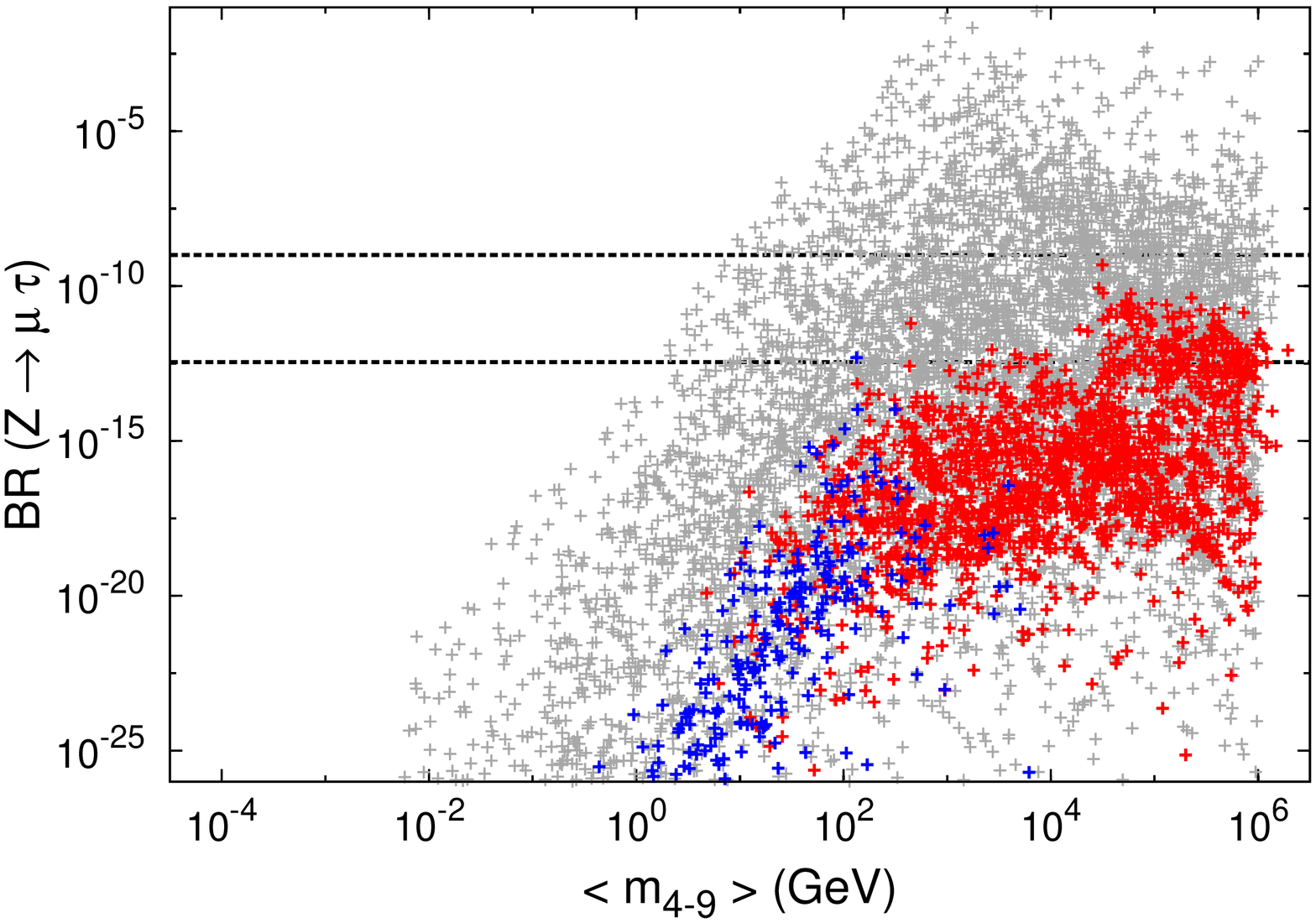, width=50mm, angle=270}
&
\epsfig{file=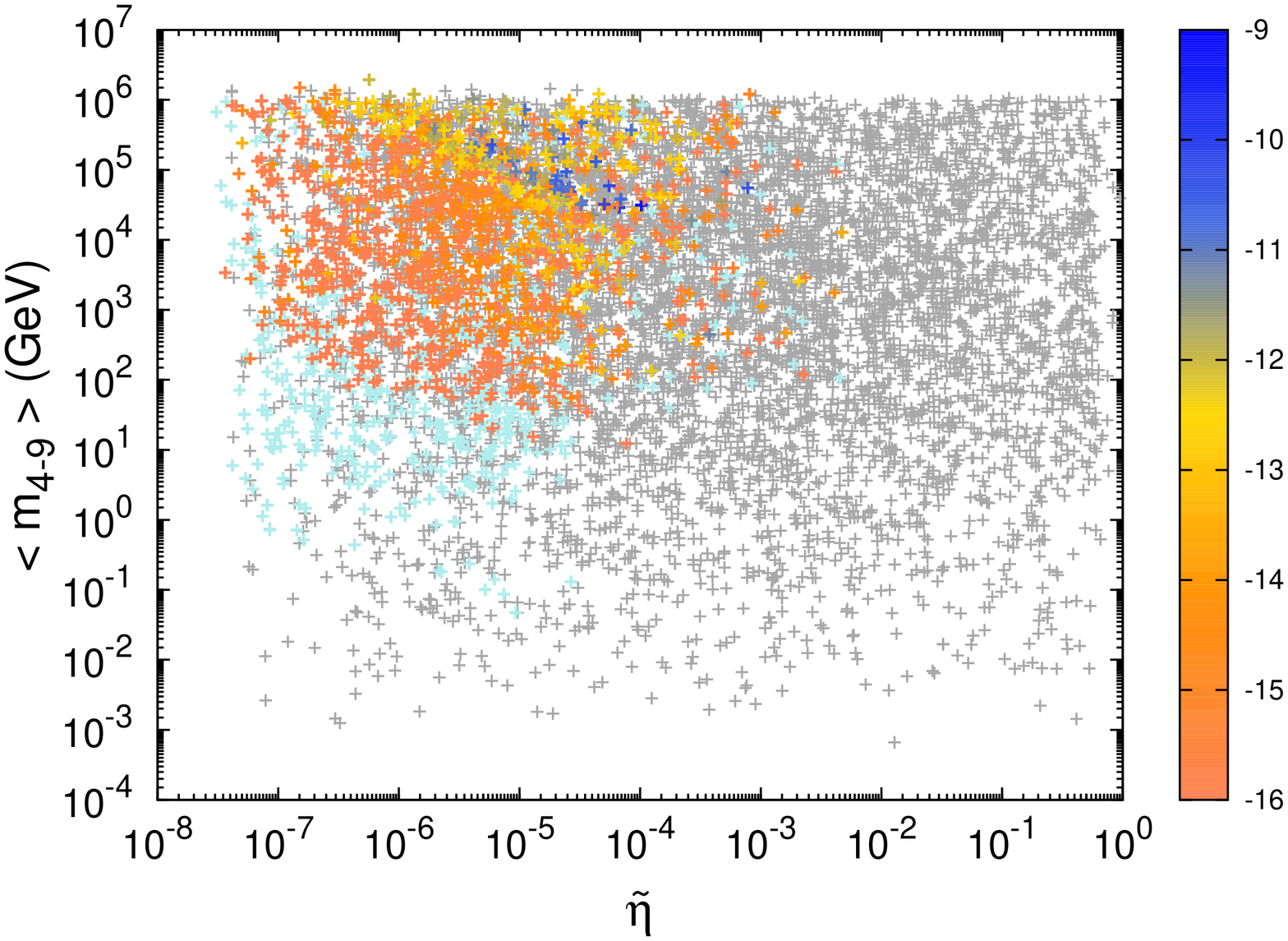, width=50mm,
  angle=270}
\end{tabular}
\caption{ISS realisation: BR($Z \to \mu \tau$) as a function of the average value of the mostly
  sterile state masses (right), $\langle m_{4-9} \rangle$ (see text for the description of the color code), 
  for a NH light neutrino
  spectrum (left);  maximal values (in log scale) of 
  BR($Z  \to \ell_1^\mp \ell_2^\pm$) 
  on the $(\tilde \eta,\langle m_{4-9} \rangle)$ 
  parameter space  (right) for a NH light neutrino spectrum, from larger 
  (dark blue) to smaller (orange) values. 
  Cyan denotes values of the branching fractions below
  $10^{-18}$. }\label{fig:ISS:BRmt.eta.avmass}
\end{figure}

\vskip -0.2cm
\noindent {\bf{cLFV $Z$ decays in the ISS}}\\
\noindent 
We show the results for this well motivated framework of neutrino mass generation 
 in Fig.~\ref{fig:ISS:BRmt.eta.avmass}.
On the left, we show the BR($Z\to \mu \tau$) as a function of the average of
the absolute masses of the mostly sterile states, $
\langle m_{4-9} \rangle\,=\, \sum_{i=4...9} \frac{1}{6}\,|m_i|$. 
In the plot on the left, we identify as grey points the solutions which fail to comply with (at least) one of the constraints listed in Section \ref{sec:constraints}.
We depict in red the points that survive all other bounds but
are typically disfavoured from standard cosmology arguments. Finally,  
blue points are in
agreement with {\it all} imposed constraints. 
These results
indicate that this ISS realisation can account for sizeable values of cLFV
$Z$-decay branching ratios, at least for the second and third generations of leptons, but mostly for cosmological disfavoured solutions. 
This in general requires
sterile states with a mass~$\gtrsim \Lambda_\text{EW}$, and can occur
even for very mild deviations from unitarity of the $\tilde
U_\text{PMNS}$. 
Other cLFV decays, $Z\to e \mu$ and  $Z\to e \tau$ 
have BRs $\lesssim \mathcal{O}(10^{-11})$, 
but still within experimental sensitivity.
The prospects for the
observation of cLFV $Z$ decays in this framework are summarised in the right plot of Fig.~\ref{fig:ISS:BRmt.eta.avmass} by
considering the values of BR($Z  \to \ell_1^\mp \ell_2^\pm$) in the $(\tilde \eta, \langle m_{4-9} \rangle)$ 
parameter space of this specific realisation and for a NH light neutrino spectrum.

\begin{figure}
\begin{center}
\begin{tabular}{cc}
\epsfig{file=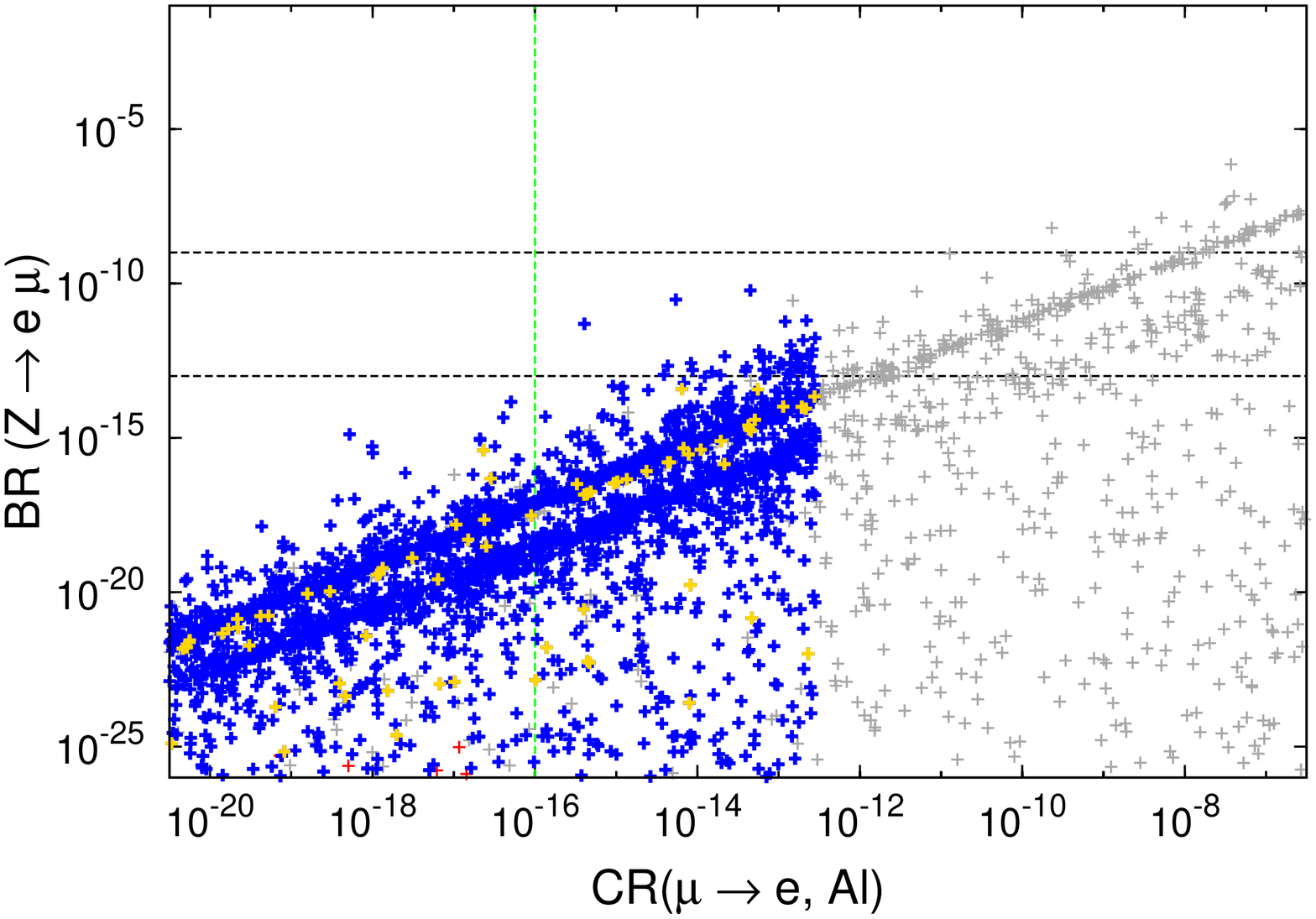, width=50mm,
  angle=270} 
 &
\epsfig{file=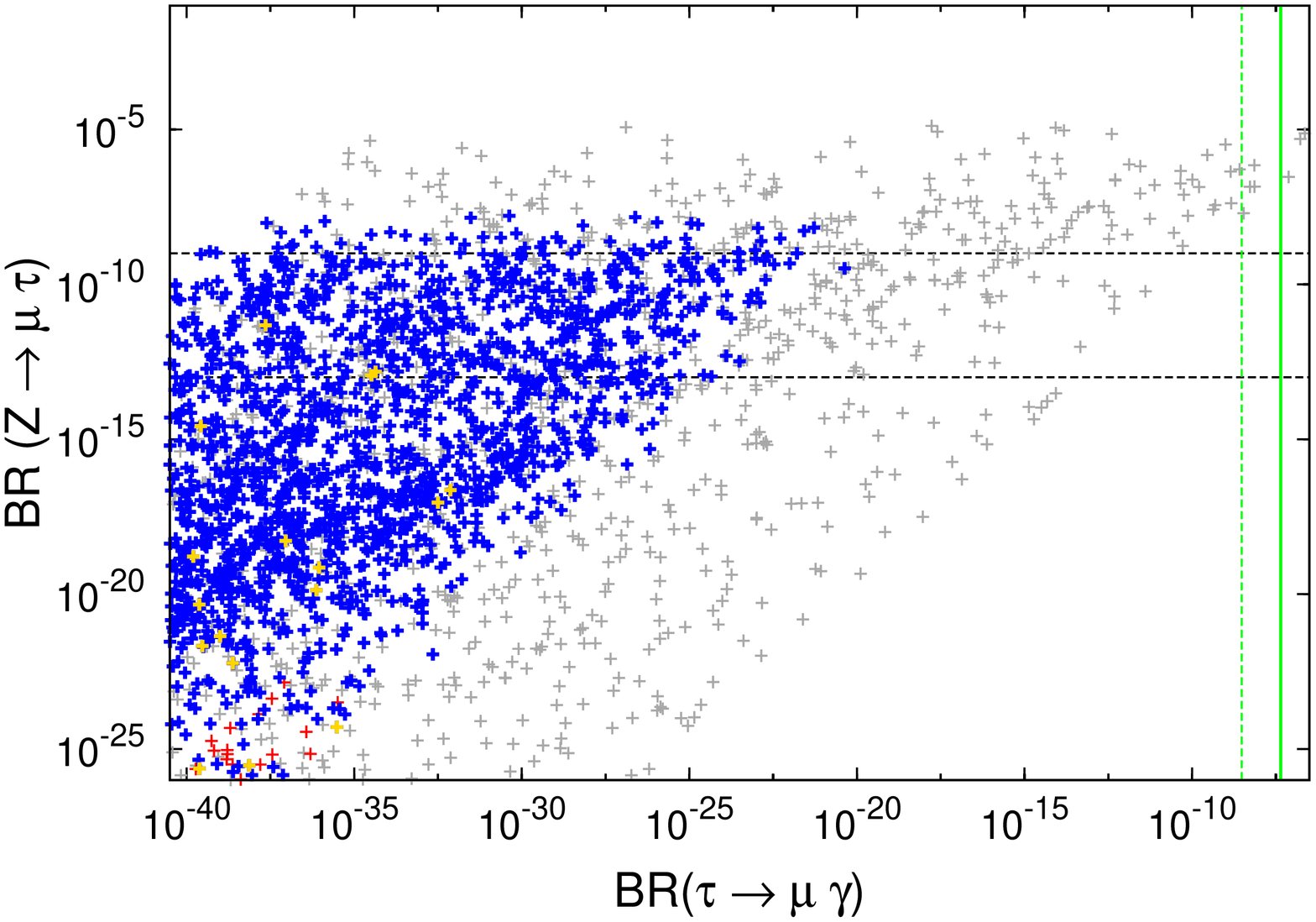,
  width=50mm, angle=270} 
\end{tabular}
\end{center}
\caption{The ``3+1 model": on the left, BR($Z \to e \mu$)
  versus CR($\mu - e$, Al), and
  BR($Z \to \mu \tau$) vs BR($\tau \to \mu \gamma$)
  on the right, for a NH light neutrino spectrum (IH leads to similar results). 
See text for the description of the color code.
  When present, 
the additional green vertical 
lines denote the current bounds (solid) and future
sensitivity (dashed), and dark-yellow points denote an associated 
$|m_{ee}|$ within experimental reach.
}\label{fig:3+1:BRZ.low}
\end{figure}

\noindent {\bf{cLFV $Z$ decays in the ``3+1 model''}}\\
\noindent 
This minimal extension of the SM by one sterile neutrino can account for values of BR($Z \to
\ell_1^\mp \ell_2^\pm$) within the sensitivity of a high luminosity
$Z$-factory, such as the FCC-ee. 
Nevertheless, the largest cLFV $Z$ decay branching
fractions (as large as $\mathcal{O}(10^{-6})$)
cannot be reconciled with current bounds on 
low-energy cLFV processes. 
Indeed, sterile neutrinos also contribute via $Z$ penguin diagrams to cLFV 3-body decays and
$\mu-e$ conversion in nuclei, which severely
constrain the flavour violating $Z \ell_1^\mp \ell_2^\pm$ vertex 
(see also~\cite{Perez:2003ad}).
Moreover, the recent MEG result on 
$\mu \to e \gamma$ also excludes important regions of the parameter space. 
These constraints are especially manifest in the case of $Z \to e \mu$ decays, since
the severe limits from BR($\mu \to 3 e$) and CR($\mu-e$, Au) typically
preclude BR$(Z \to e \mu) \gtrsim 10^{-13}$.
In Fig.~\ref{fig:3+1:BRZ.low} we illustrate the complementary r\^ole of a high-luminosity $Z$-factory
with respect to low-energy (high-intensity) cLFV dedicated
experiments (same color code as in Fig.~\ref{fig:ISS:BRmt.eta.avmass}).  We display the sterile neutrino contributions to BR($Z \to \ell_1^\mp
\ell_2^\pm$) versus two different low-energy cLFV observables:  CR($\mu - e$, Al) and
 BR($\tau \to \mu \gamma$). We further highlight in dark yellow solutions which allow for a third complementary observable within 
future sensitivity, which is the effective neutrino mass in
$0\nu 2 \beta$ decays.  
 \vskip -1.0cm
\noindent

As can be inferred from
Fig.~\ref{fig:3+1:BRZ.low}, low-energy cLFV dedicated facilities offer
much better prospects to probe LFV in the
$\mu-e$ sector of the ``3+1 model'' than a high-luminosity $Z$-factory. In
particular, Mu3e (PSI) \cite{Blondel:2013ia} and COMET (J-PARC)
\cite{Kuno:2013mha} will be sensitive to regions 
in parameter space associated with BR($Z \to e \mu$)~$\sim 10^{-17\div
  -13}$, beyond the reach of FCC-ee. Interestingly, the situation is
reversed for the case of the $\mu-\tau$ sector. Moreover, a non negligible subset of the parameter space is testable at a third type of facilities, through  $0\nu 2 \beta$  decay searches (especially in the
case of an IH light neutrino spectrum, although we have not displayed it here).

\vskip -0.6cm
\section{Conclusions}
\vskip -0.2cm
\noindent
We have considered two extensions of the SM which add to its particle content one or more sterile neutrinos. We have explored indirect searches for these sterile states at a future circular collider like FCC-ee, running close to the $Z$ mass threshold.
We have considered the contribution of the sterile states to rare cLFV $Z$ decays in these two classes of models and discussed them taking into account a number of experimental and theoretical constraints.
Among these, low-energy LFV observables like cLFV 3-body decays and
$\mu-e$ conversion in nuclei impose strong constraints on the sterile neutrino induced BR($Z  \to \ell_1^\mp \ell_2^\pm$).
Our analysis emphasises the underlying synergy between a high-luminosity $Z$ factory and dedicated low-energy facilities: regions of the parameter space of both models can be probed via LFV $Z$ decays at FCC-ee, at low-energy cLFV dedicated facilities and also via searches for $0\nu 2 \beta$. 
Notably, FCC-ee could better probe LFV in the $\mu-\tau$ sector, in complementarity to the reach of low-energy experiments like COMET.
\vskip 0.2cm
\section*{Acknowledgments}
\vskip -0.2cm 
\noindent
We acknowledge support from the EU FP7 ITN
INVISIBLES (Marie Curie Actions, PITN-\-GA-\-2011-\-289442).


\begin{thebibliography}{99}



\bibitem{Abazajian:2012ys}
  K.~N.~Abazajian {\it et al.}, 
  arXiv:1204.5379 [hep-ph].
   
  \bibitem{Kusenko:2009up}
  A.~Kusenko,
  Phys.\ Rept.\  {\bf 481} (2009) 1
  [arXiv:0906.2968 [hep-ph]].
  
  \bibitem{Gomez-Ceballos:2013zzn}
M. Bicer {\it et al.}, 
JHEP  {\bf 1401} (2014) 164 
[arXiv:1308.6176 [hep-ex]].
  
\bibitem{Abada:2014cca}
  A.~Abada {\it et al.},
  JHEP {\bf 1504} (2015) 051
  [arXiv:1412.6322 [hep-ph]].
  
  \bibitem{Glashow:1970st}
S.~L. Glashow {\it et al.}, {Phys. Rev. D} {\bf 2} (1970)
  1285.
  
  \bibitem{LFVZdecays}
  G.~Mann and T.~Riemann,
  Annalen Phys.\  {\bf 40} (1984) 334; 
  J.~I.~Illana {\it et al.},
  in {\it 2nd ECFA/DESY Study 1998-2001}, 490-524
  [hep-ph/0001273]; 
  J.~I.~Illana and T.~Riemann,
  Phys.\ Rev.\ D {\bf 63} (2001) 053004
  [hep-ph/0010193]; 
  A.~Ilakovac and A.~Pilaftsis,
  Nucl.\ Phys.\ B {\bf 437} (1995) 491
  [hep-ph/9403398]; 
  S.~Davidson, S.~Lacroix and P.~Verdier,
  JHEP {\bf 1209} (2012) 092
  [arXiv:1207.4894 [hep-ph]].

  \bibitem{Perez:2003ad}
  M.~A.~Perez {\it et al.},
  Int.\ J.\ Mod.\ Phys.\ A {\bf 19} (2004) 159
  [hep-ph/0305227];   
  A.~Flores-Tlalpa {\it et al.},
  Phys.\ Rev.\ D {\bf 65} (2002) 073010
  [hep-ph/0112065]; 
  D.~Delepine and F.~Vissani,
  Phys.\ Lett.\ B {\bf 522} (2001) 95
  [hep-ph/0106287].
  


  \bibitem{Mohapatra:1986bd}
  R.~N.~Mohapatra and J.~W.~F.~Valle,
  Phys.\ Rev.\ D {\bf 34} (1986) 1642.


    \bibitem{GonzalezGarcia:1988rw} 
  M.~C.~Gonzalez-Garcia and J.~W.~F.~Valle,
  Phys.\ Lett.\ B {\bf 216} (1989) 360.
  
  \bibitem{FernandezMartinez:2007ms}
  E.~Fernandez-Martinez {\it et al.},
  Phys.\ Lett.\ B {\bf 649} (2007) 427
  [hep-ph/0703098].


\bibitem{Schechter:1980gr}
  J.~Schechter and J.~W.~F.~Valle,
  Phys.\ Rev.\ D {\bf 22} (1980) 2227.

\bibitem{Gronau:1984ct}
  M.~Gronau, C.~N.~Leung and J.~L.~Rosner,
  Phys.\ Rev.\ D {\bf 29} (1984) 2539.
  
  

  
  \bibitem{Forero:2014bxa}
  D.~V.~Forero {\it et al.},
  Phys.\ Rev.\ D {\bf 90} (2014) 093006
  [arXiv:1405.7540 [hep-ph]].


\bibitem{Antusch:2008tz} 
S.~Antusch {\it et al.},
Nucl.\ Phys.\ B {\bf 810} (2009) 369 
[arXiv:0807.1003 [hep-ph]].


\bibitem{Antusch:2014woa}
  S.~Antusch and O.~Fischer,
  JHEP {\bf 1410} (2014) 94
  [arXiv:1407.6607 [hep-ph]].

\bibitem{Blondel:2013ia}
  A.~Blondel  {\it et al.},
  arXiv:1301.6113 [physics.ins-det].
  
     \bibitem{Kuno:2013mha}
  Y.~Kuno [COMET Collaboration],
  PTEP {\bf 2013} (2013) 022C01.
  
  

\end{thebibliography}
\end{document}